\documentclass[a4paper,11pt]{iopart}

\usepackage{xcolor}
\usepackage{graphicx}
\usepackage{caption}
\usepackage{iopams}

\usepackage{hyperref}   
 \hypersetup{
    colorlinks = true,
    linkcolor = red,
    citecolor = blue,
}

\newcommand{\sectionred}[1]{\section*{\large\textcolor{red}{#1}}}

\begin{document}

\title{Characterization methods dedicated to nanometer-thick hBN layers}

\author{L\'eonard Schu\'e$^{1,2}$, Ingrid Stenger$^{2}$, Fr\'ed\'eric Fossard$^{1}$, Annick Loiseau$^{1}$ and Julien Barjon$^{2}$}
\address{$^1$ Laboratoire d'\'Etude des Microstructures (LEM), CNRS, ONERA, Universit\'e Paris-Saclay, Ch\^atillon, France}
\address{$^2$ Groupe d'\'Etude de la Mati\`ere Condens\'ee (GEMaC), CNRS, UVSQ, Universit\'e Paris-Saclay, Versailles, France}
\ead{julien.barjon@uvsq.fr; annick.loiseau@onera.fr}

\begin{abstract}

\noindent
Hexagonal boron nitride (hBN) regains interest as a strategic component in graphene engineering and in van der Waals heterostructures built with two dimensional materials. It is crucial then, to handle reliable characterization techniques capable to assess the quality of structural and electronic properties of the hBN material used. We present here characterization procedures based on optical spectroscopies, namely cathodoluminescence and Raman, with the additional support of structural analysis conducted by transmission electron microscopy. We show the capability of optical spectroscopies to investigate and benchmark the optical and structural properties of various hBN thin layers sources.
 
\end{abstract}

\maketitle

\sectionred{Introduction}

Two-dimensional (2D) crystals are receiving considerable attention due to their new electronic and optical properties compared to their bulk counterpart over a large range of the electromagnetic spectrum, thus opening up new perspectives for future nanoelectronic or optoelectronic applications \cite{Novoselov2004,Novoselov2005,Radisavljevic2011,Xia2014}.

Unlike graphene, these materials are semiconductors with optical properties dominated by pronounced excitonic effects which are currently intensively studied in transition-metal dichalcogenides (TMDs) \cite{Zhao2012,Xu2013,Zhang2015a,Molina2013,Wu2015a}, phosphorene \cite{Ling2015a,Favron2015} and hexagonal boron nitride (h-BN) \cite{Watanabe2004,Jaffrennou2007,Watanabe2009,Museur2011,Galambosi2011,Pierret2014,Schue2016,Fugallo2015,Galvani2016}. This latter material is a particular case since it is a large bandgap ($>$6eV) material with a honeycomb lattice similar to that of graphene where boron and nitrogen atoms alternate at the vertices of the planar hexagonal sp$^{2}$ network. In recent years, this material has attracted growing attention appearing as an ideal insulating support of graphene in electronic devices or encapsulating medium of 2D layers. This ability is mainly due to its unique physical properties such as atomic flatness, chemical inertness, absence of dangling bonds or low dielectric screening \cite{Xue2011,Liu2013}. As a consequence, two-dimensional h-BN layers have become an incontournable component in graphene engineering \cite{Dean2010,Ramasubramaniam2011,Yankowitz2012,Yang2013} as well as in the realization of van der Waals 2D heterostructures \cite{Britnell2012,Withers2015,Cao2015}. On the other hand, several studies have also demonstrated its high radiative efficiency close to 6 eV offering new optic emission capabilities in the ultraviolet \cite{Watanabe2004,Kubota2007,Watanabe2009b}.

Today these achievements are based on the use of mechanically exfoliated layers mostly from single crystals produced at NIMS \cite{Taniguchi2007}. To go beyond the limitations of exfoliation procedures, efficient and scalable routes are searched to the routine synthesis of large BN layers and films, which remains a challenging task \cite{Hemmi2014,Jang2016,Yuan2016,Wu2015b,Kim2015}. A second challenge is to handle reliable tools for characterizing and studying the electronic and optical properties of the h-BN material both as bulk and as a function of the number of layers. This task is indeed difficult because of the necessity to work in the far UV range, an energy range accessible in only a few experimental set-ups. Nevertheless, this knowledge is essential for assessing the quality of a given h-BN source. The nature and the density of defects it may contain can also affect its own properties or its electronic coupling with other 2D materials. For instance, it has been recently predicted that encapsulating MoS$_{2}$ and Black Phosphorus layers into hBN layers may significantly change their electronic/excitonic properties due to screening effects \cite{Haastrup2016,Steinkasserer2016}.

In this paper, we present characterization procedures capable to address these issues and to investigate the optical, structural and electronic properties of hBN layers. We first describe our current routine method to accurately determine the layer thickness, which becomes tricky for thin flakes below 20 layers (20L). We then investigate and compare the potentialities of two optical spectroscopies, Raman and cathodoluminescence, with the additional support of TEM structural analyses. We show that cathodoluminescence provides a powerful investigation tool to evaluate the hBN purity and crystallinity through a comparative study of BN sources obtained by various growth processes.

\sectionred{Results and Discussions}

\subsection*{\textbf{Thickness measurements}}

\qquad A critical step in 2D material study is the accurate determination of their thickness. Various approaches are currently used for thin hBN samples (in the 1-20L range) transferred either on SiO$_2$/Si substrate or on TEM grids.

\begin{figure}[ht]
 \centering
\includegraphics[scale=0.81]{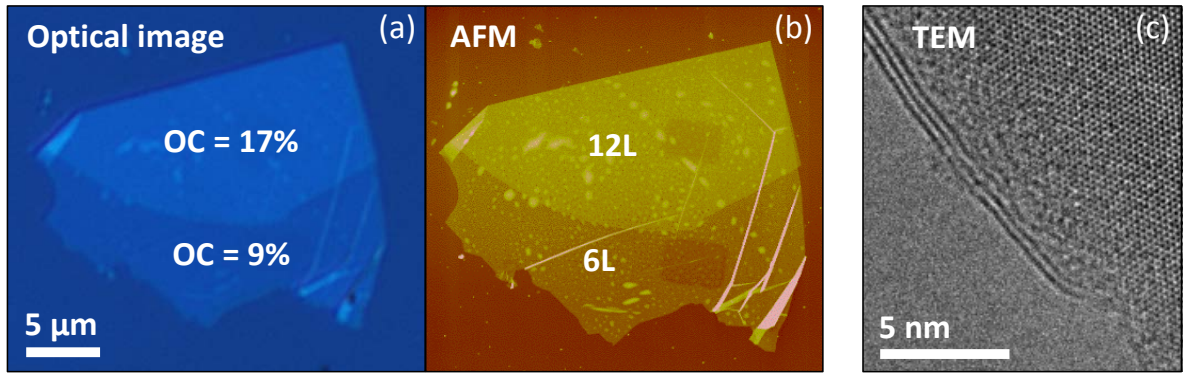}
\caption{(a) Optical and (b) AFM images of a 6L hBN folded flake. (c) TEM image of a bi- and trilayer folded flake.}
\label{F1}
 \end{figure}

Nanometer-thin hBN films deposited onto SiO$_2$/Si substrate are usually first examined by using optical simple microscopy techniques. A preliminary identification is conducted to rapidly localize the thinnest areas. The role of the SiO$_2$ layer thickness is important and well documented in graphene: with the suited oxide thickness, a monolayer graphene presents an optical contrast (OC) up to 10\% in the visible range \cite{Blake2007} allowing us to see an atomic layer of carbon with eyes. For hBN, the best OC that can be attained does not exceed 2.5\% per layer when using the optimal SiO$_2$ thickness (80$\pm$10 nm or 290$\pm$10 nm) \cite{Gorbachev2011,Golla2013}. This makes the OC approach more challenging for hBN than for graphene. The OC approach is also excluded for self-standing layers as for layers deposited on other substrates than SiO$_2$/Si. Other techniques are often needed.

Atomic Force Microscopy (AFM) is widely employed but confronted to various obstacles in the low thickness range especially (1-20L). First, surface roughness of the substrate can be really troublesome for very thin samples but also surface contaminations like adhesive residue (due to exfoliation process) or exogenous impurities introduced during CVD film transfer for instance. Moreover, a thin water layer captured between the substrate and the hBN is often observed due to the hydrophilic character of SiO$_2$ as discussed in a previous study \cite{Pierret2014}. At the end, AFM gives an uncertainty on thickness usually higher than $\pm$1L. A possibility to circumvent the water layer issue consists in looking at folded hBN flakes like the one depicted in figure \ref{F1}(a-b). Such a configuration ensures a more accurate estimation of the AFM step height between the folded part and the unfolded one, independently on the water layer trapped at the hBN/SiO$_2$ interface. Then the uncertainty of the folded hBN layer thickness is about $\pm$0.1L. Besides, we reported a linear evolution of the OC with the number of hBN layers \cite{Schue2016}. Once calibrated on the basis of folded flakes, the OC allows to measure the thickness of any other flakes (folded or not) on the same wafer with an error of about $\pm$0.2L in the 1-20L range.

When transferred on TEM grids, a direct strategy for measuring the hBN layer thickness relies likewise on folded flakes \cite{Han2008}. Indeed, the fringe contrast visible on figure \ref{F1}(c) along the fold axis leads to count the number of hBN layers, directly. Contaminants or external pollutions clearly observable on the right part of figure \ref{F1}(c) do not impede measurement in this case. Other studies have reported the use of sample tilt looking at diffraction spots intensity \cite{Pan2012} or have also exploited reconstructed phase images for one- to four-layers BN identification in HRTEM imaging \cite{Alem2009}.

\begin{figure}[ht]
 \centering
\includegraphics[scale=1.6]{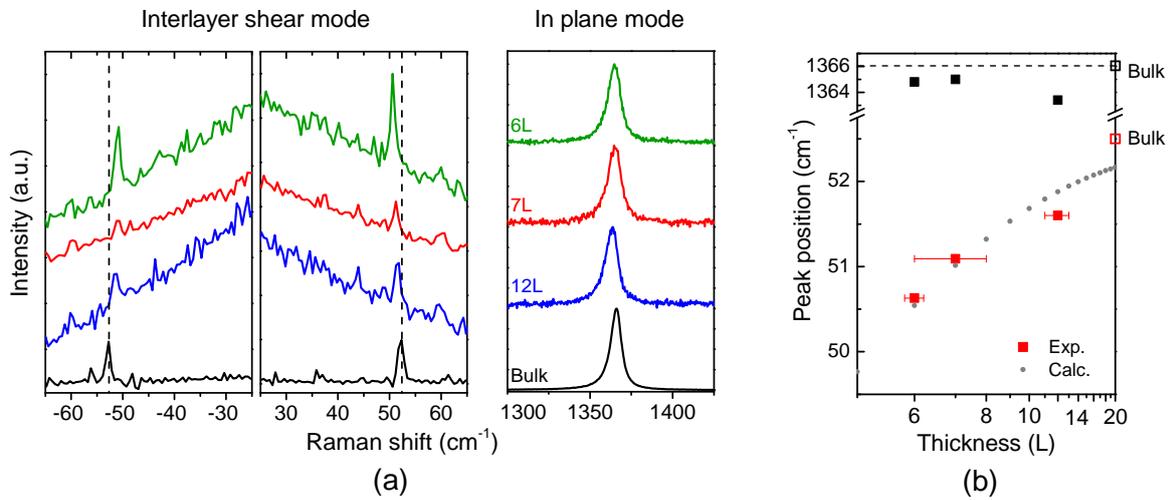}
\caption{(a) Stokes and Anti-Stokes low-frequency (left) and high frequency (right) Raman spectra of 6L, 7L, 12L exfoliated on SiO$_{2}$/Si substrate and bulk hBN samples. (b) Energy position of the interlayer shear mode (red square) and of the in plane mode (1366 cm$^{-1}$, black square) as a function of the number of hBN layer.}
\label{F2}
 \end{figure}

More recently, Tan and coworkers \cite{Tan2012} have put forward a novel approach to determine the number of graphene layers by Ultra-Low Frequency (ULF) Raman spectroscopy. It is based on a vibrational mode characteristic of the 2D materials: an interlayer shear mode associated to the relative motion of adjacent atomic planes. It emerges at low-frequency and strongly depends on the layer number. Its frequency is found to decrease from 43 cm$^{-1}$ in bulk graphite down to 31 cm$^{-1}$ in a graphene bilayer (no shear mode for a monolayer) while only the bulk value has been measured experimentally for hBN close to 52.5 cm$^{-1}$ \cite{Kuzuba1978}. It is important to notice that this shear mode presents discrete frequency values for each thickness. Then ULF Raman spectroscopy appears very promising for measuring the number of hBN layers with an uncertainty lower than a single monolayer.

This procedure can then be extended to other 2D materials as already reported for MoS$_{2}$ \cite{Zeng2012,Zhang2013} and Black Phosphorus \cite{Ling2015b} and might be suitable for all kind of samples (on substrate or self-suspended). We have investigated the ULF approach on thin hBN samples as shown in figure \ref{F2}. Theoretical values of this low-frequency mode have been calculated following the same simple model (finite linear chain) used for graphene (gray dots figure \ref{F2}(b)). Nevertheless, contrary to the graphene case, Raman processes are \textit{non-resonant} when exciting with visible laser sources because of its large bandgap ($>$6 eV). As a consequence, the ULF Raman signal from nanometer-thin hBN layers is much weaker than in other 2D materials. It requires long integration times, high laser power excitation and special care to minimize the noise level. So far, we investigated the bulk material and thin layers down to 6L. Stokes and Anti-Stokes Raman spectra presented in figure \ref{F2}(a) display the low-frequency peaks emerging at 52.5, 51.6, 51.1 and 50.6 cm$^{-1}$ recorded on bulk hBN and flakes composed of 12L, 7L and 6L respectively. The downshift of the ULF shear mode is observed in hBN of a few atomic layer thickness, almost following the theoretical model used for other 2D materials as shown in figure \ref{F2}(b). Note that the thickness error bar is sometimes large ($\pm$1L) and corresponds to situations where folded flakes could not be found on the wafer as discussed previously. 

In parallel, the Raman signal of the high frequency mode, with a nominal value at 1366 cm$^{-1}$, was simultaneously recorded (figure \ref{F2}(a)) and is plotted in figure \ref{F2}(b) for each sample. This in plane vibrational mode is extensively used to assess the sample crystallinity as discussed in the next part. For thin hBN layers, we found lower frequencies than the one expected for the bulk at 1366 cm$^{-1}$. Since we excite the samples with a relatively high laser power, we suspect a local heating to be responsible of this frequency shift. This actually can also affect the low frequency mode. We are still currently working on refining the experimental methodology. Temperature effects have been examined and will be the subject of a forthcoming paper.

\subsection*{\textbf{Structural properties from Raman spectroscopy}}

\begin{figure}[ht]
 \centering
\includegraphics[scale=0.7]{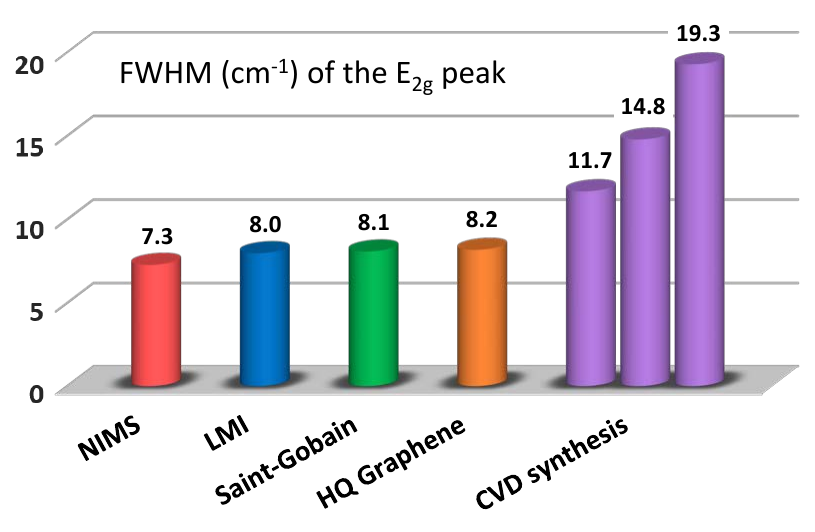}
\caption{FWHM of the Raman high frequency peak (E$_{\rm{2g}}$ mode) measured on four hBN sources (NIMS, LMI, Saint-Gobain and HQ samples) compared to three values taken from the literature on CVD thin films \cite{Song2010,Lee2012,Tay2014}.}
\label{F3}
 \end{figure}

\qquad Raman spectroscopy is the non-destructive technique widely used in the 2D material community. It is routinely employed to provide a rapid feedback in the optimization of CVD growth processes of hBN for instance. The Raman peak which is conventionally investigated is observed at 1366 cm$^{-1}$. It is related to the in-plane atom vibrations (E$_{\rm{2g}}$ mode) \cite{Geick1966} analogous to the so-called G peak in graphene. However, owing to its non-resonant character the main limitation of conventional Raman spectroscopy for hBN lies in the scarcity of information which can be drawn from a single peak. Its FWHM remains, up to now, the most widely used characterization of the hBN crystallinity. Values measured on four different hBN bulk sources are presented in figure \ref{F3} together with other CVD data taken from the literature \cite{Song2010,Lee2012,Tay2014}.

The comparison of Raman FWHM then reveals that CVD films are of significantly lower quality than the HPHT, polymer-derived growth, or commercial powders. The latter are found within the 7-8 cm$^{-1}$ FWHM range in the same experimental conditions for the bulk material and below 10 cm$^{-1}$ after exfoliation. These measurements are therefore not discriminating the quality of these four BN sources.

\subsection*{\textbf{Luminescence spectroscopy}}

\subsubsection*{\qquad a - General luminescence features of hBN\\\\}
 
\qquad While the electronic properties of graphene or TMDs are well-described theoretically and thoroughly investigated by means of conventional spectroscopies, the ones of hBN are less known. Widely used for semiconductor analysis, luminescence spectroscopies appear to be a powerful tool for exploring the hBN physical properties. A typical cathodoluminescence (CL) spectrum of hBN is displayed in figure \ref{F4} in the 200-1000 nm wavelength range.

Cathodoluminescence images, shown below, were recorded on a thin lamellae prepared by focused ion beam (FIB). The slab which is studied in figure \ref{F7} was cut along the [00.1] direction from the HPHT crystal for plane view observations. It will be further examined through a TEM analysis in the last part of the manuscript (see figure \ref{F7}).

\begin{figure}[ht]
 \centering
\includegraphics[scale=0.5]{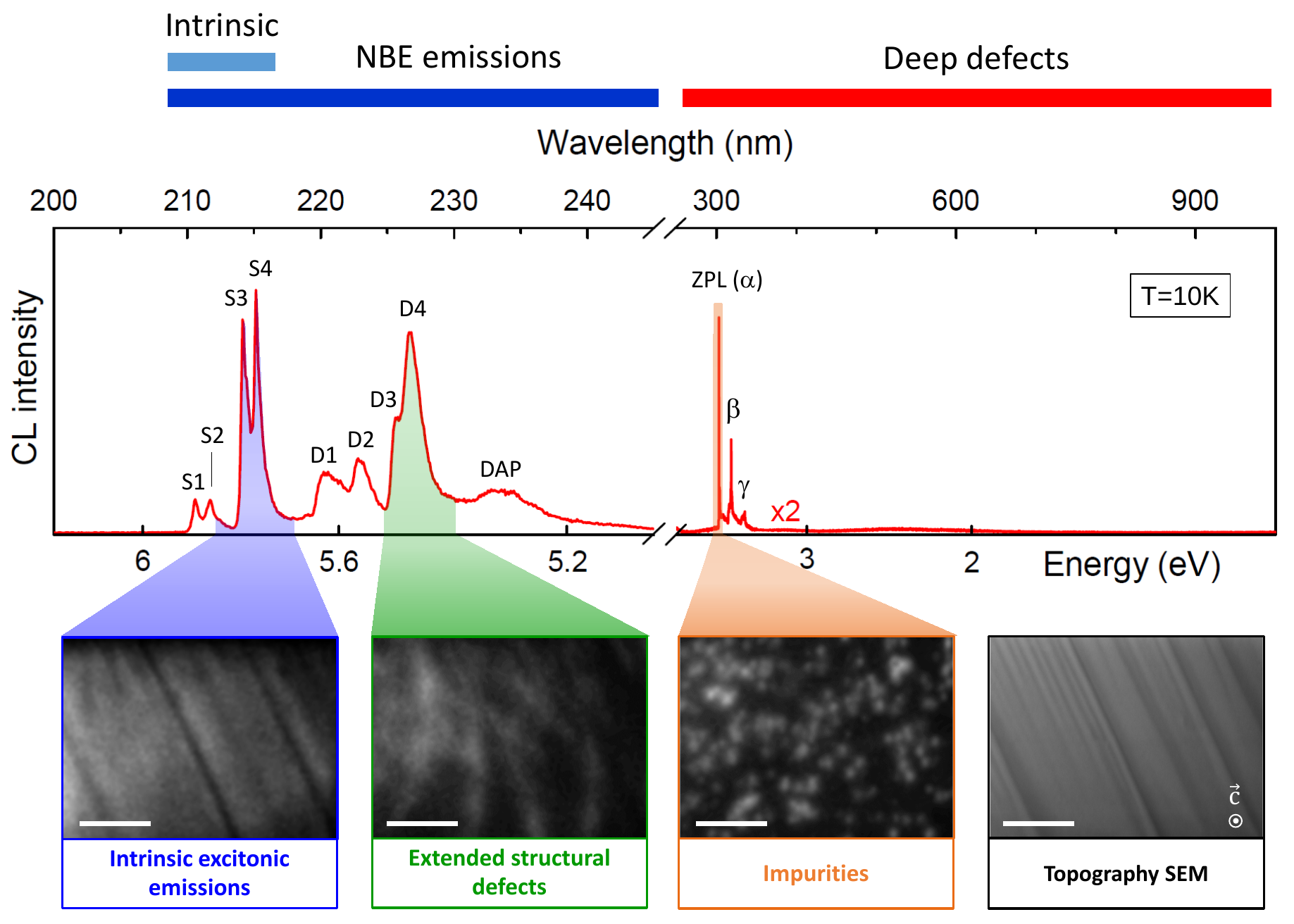}
\caption{Upper: Typical cathodoluminescence spectrum of hBN in the 200-1000 nm range acquired at 10K on a FIB preparation of the HPHT (see additional analysis by TEM in figure \ref{F7}). Lower: SEM image (black) with the corresponding monochromatic CL images recorded in the same area at $\pm$3 nm around 215 nm (blue), 227 nm (green) and 303 nm (orange).
Scale bar: 1$\mu$m.}
\label{F4}
 \end{figure}

In the next paragraphs, the CL spectrum is described starting from the shortest to the largest wavelengths.
Luminescence properties of hBN are governed by unusually strong excitonic effects resulting in recombinations at energies below the bandgap energy. These are the so-called near band edge (NBE) recombinations located in the deepest UV region presented in figure \ref{F4}. Among them, a series of peaks at the shortest wavelength (210-215 nm) is referred as the S series which intrinsic origin was proven recently in an investigation of hBN materials obtained from different growth procedures \cite{Schue2016}. The exciton nature of these recombinations is well established theoretically \cite{Blase1995,Wirtz2005,Arnaud2006,Galvani2016} but their fine structure splitting and their correlation with absorption data are still under debate \cite{Schue2016,Watanabe2009,Museur2011,Cao2013,Bourrellier2014,Li2016b,Cassabois2016,Doan2016}.

Other luminescence features can be observed in the 210-250 nm region tentatively assigned here to the NBE region. At higher wavelengths than the S series, the most documented luminescence feature is the so-called D series (220-227 nm), see figure \ref{F4} related to the presence of extended structural defects. Its origin is not completely elucidated yet but it has been experimentally correlated to the presence of stacking defects \cite{Watanabe2006,Watanabe2006b,Watanabe2011b}, dislocations \cite{Jaffrennou2007} or grain boundaries \cite{Pierret2014}. These emissions arise from recombinations of excitons trapped on structural defects. Note the ambiguity between the attribution of D1 and the first phonon replica of S4 emerging at the same wavelength \cite{Pierret2014}. Finally, a weak and broader emission is often observed at 233 nm, such as in figure \ref{F4}, and has been attributed to donor-acceptor pairs (DAP) recombination processes \cite{Museur2008,Du2016}.

At higher wavelengths ($>$250 nm), the luminescence spectrum reveals deep defect emissions. We voluntarily present a large wavelength range till 1000 nm in order to evidence the absence of significant signal above 400 nm. A luminescence signal with a maximum at 302.8 nm is regularly detected in various hBN sources (see also figure \ref{F5}). This emission is extremely sharp: its linewidth is lower than 0.075 nm, the spectral resolution of the experiment. It has been identified as the zero-phonon emission line (ZPL) labelled $\alpha$ of substitutional impurities introduced in the lattice during the synthesis, probably carbon (or possibly oxygen) as revealed by a detailed SIMS analysis \cite{Taniguchi2007}. The $\beta$ and $\gamma$ peaks were assigned to replicas, with an energy of 195 meV, involving a local vibrational mode (LVM) \cite{Silly2007,Museur2008b} or phonons of the hBN bulk crystal lattice \cite{Vuong2016}. In recent studies, similar additional sharp emissions have been reported in hBN samples in the visible range after introducing intentionally defects in the crystal \cite{Tran2016}.

CL monochromatic images confirm the distinct origins of these luminescence features. They were taken on the plane-view FIB sample at three distinct wavelengths corresponding to the three regions introduced before: intrinsic, near band-edge and deep defects. The image taken from the intrinsic emissions (S series) was obtained by collecting photons emitted at 215$\pm$3 nm and is almost homogeneous except along on parallel black lines. This presumably comes from thickness variations along the thin slab. Typical "curtaining effects"€ from the ion bombardment during FIB preparation are indeed observed in secondary electron images. The CL image taken from the D series (227$\pm$3 nm) displays several continuous lines crossing the sample in random directions. They are clearly not correlated to the intrinsic emissions. The third CL image of chemical impurities was recorded on the ZPL emission at 302.8$\pm$3 nm. Here again, we found a totally different spatial distribution of the luminescence. The distinctive luminescence spots observed unveil a discrete and random distribution of the impurity centers. Similar spots were recently observed and have been identified as single photon emitters, indicating that they correspond to isolated atomic centers randomly diluted in the crystal lattice \cite{Bourrellier2016}.\\

\subsubsection*{\qquad b - Comparative study of different hBN bulk materials\\\\}

\begin{figure}[ht]
 \centering
	\includegraphics[scale=1.92]{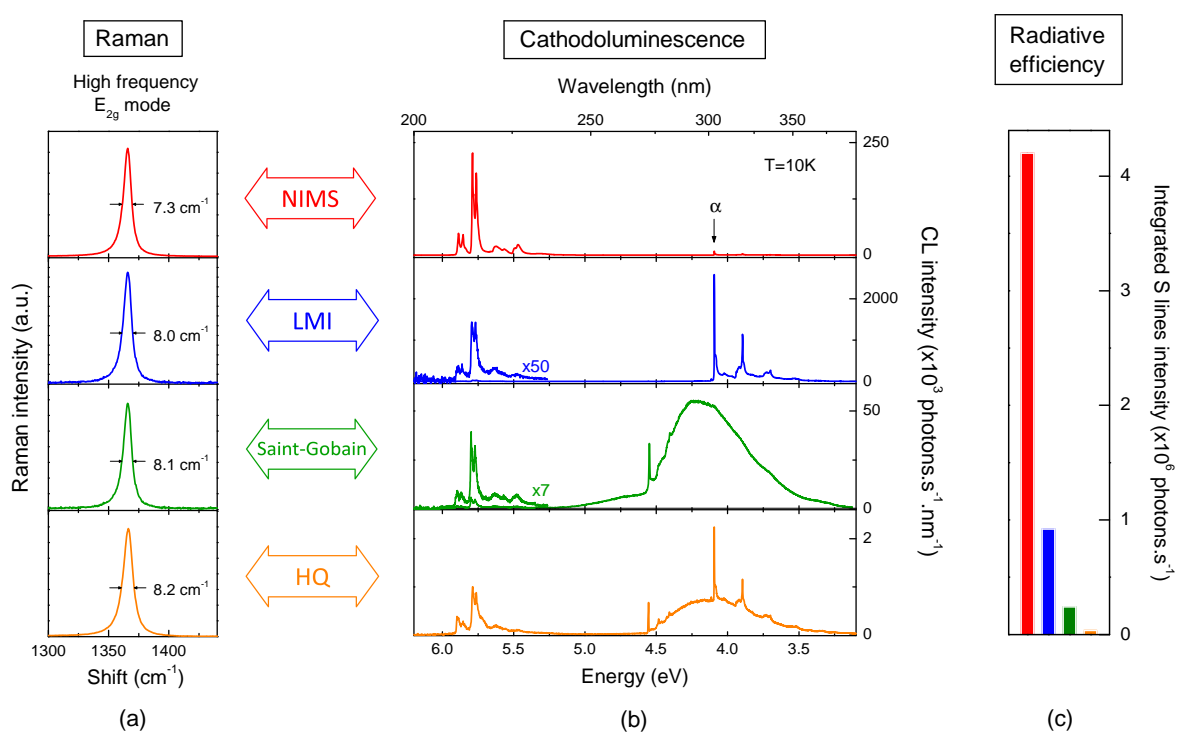}
	\caption{Typical (a) Raman and (b) CL spectra obtained for the NIMS (red), LMI (blue), Saint-Gobain (green) and HQ (orange) hBN sources. All CL spectra were acquired at 10K under 2 keV-1 nA electron beam excitation. Spectra are corrected from the spectral response of the detection system. (c) The recombination intensity of excitons (S lines) integrated in the 209-217 nm range.}
	\label{F5}
\end{figure}

\qquad Having previously shown that the typical luminescence spectrum of hBN offers a wealth of information, we present now a comparative study of four bulk hBN sources. We display in figure \ref{F5} both Raman and CL spectra recorded on each hBN source in the same manner done by Tsuda \textit{et al.} \cite{Tsuda2007}, but here in order to emphasize the differences between the two kinds of optical analyses. We point out that each spectrum is representative of the signal obtained for each sample. In figure \ref{F5}, it appears remarkable that for almost similar linewidths obtained from Raman spectroscopy, the luminescence spectra exhibit much diverse features depending on the hBN source.

First, one can see that the CL spectrum of the HPHT-grown hBN exhibits very intense intrinsic emissions whilst LMI, Saint-Gobain and HQ luminescence spectra are dominated by deep defects emissions. Indeed the high intensities of intrinsic emissions associated to weak deep-defects signal of the HPHT sample are in good agreement with the low FWHM values measured in Raman spectroscopy.

For high excitation energies (above the bandgap) such as the energetic electrons used in CL, injected electrons and holes accumulate at the bandgap extrema, sometimes bounded as excitons. In hBN the intrinsic luminescence is the result of radiative recombinations of compact excitons with a characteristic radiative lifetime $\tau_{\rm{rad}}$. Under continuous excitation the steady state population of excitons is equal to $G\tau$ where $G$ is the generation factor \cite{Davidson1977}, i.e. the number of excitons produced per incident electron, and $\tau$ the total lifetime carriers. So that the luminescence efficiency: $I_{CL} \propto G \tau / \tau_{\rm{rad}}$ with $I_{CL}$ the luminescence intensity.

The total exciton lifetime $\tau$ is the important parameter here (the radiative lifetime being a constant parameter determined by quantum mechanics). One has to keep in mind that many other recombination pathways are offered to the exciton population accumulated at the energy bottleneck formed at its ground state. All these pathways compete together, so that the total exciton lifetime is governed by the shorter process according to: $1 / \tau = 1 / \tau_{\rm{rad}} + \Sigma (1 / \tau_{i})$ where $\tau_{i}$ are the characteristic times of all processes other than radiative exciton annihilation.

During the experiments on bulk materials reported in figure \ref{F5}(b), the $G$ factor was kept constant. Hence the integrated intensity of intrinsic luminescence (in the 209-217 nm range) remains proportional to the $\tau/\tau_{\rm{rad}}$ ratio also known as the luminescence efficiency or internal quantum yield. Without going into details in the complex calibration procedures needed for its quantitative determination, a comparative analysis can be achieved when samples are excited in the same conditions (acceleration voltage, beam current and temperature).

For an ideal and pure crystal, i.e. without any competing recombination channel, the luminescence intensity is expected to reach the maximum 100\% quantum yield. Still, in a real crystal, the presence of defects (impurities, atomic complexes, point, plane or extended structural defects, surface\ldots) offers alternative recombination pathways - radiative or not - than the radiative exciton recombination. This is the main reason why low crystallinity and/or impure crystals exhibit lower intrinsic luminescence intensity. The comparison of NIMS and LMI samples spectra in figure \ref{F5}(b) offers a good illustration of how the impurity content can affect the intrinsic exciton luminescence. The deep-defect emissions observed in the CL spectrum (around 4 eV) of the LMI sample are accompanied by a 4 times decrease of the free exciton recombination intensity. More generally, the recombination intensity of S excitons is much more sensitive than the conventional Raman spectroscopy when assessing crystallinity and purity of hBN. While the Raman FWMH displays a 12\% variation close to the spectral resolution, the recombination intensity of S excitons varies by 2 orders of magnitude among the four specimen investigated in this work (figure \ref{F5}(c)).

As a final remark, note that the full CL spectrum, including luminescence of deep defect centers, provides complementary information. The impurity center with a maximum intensity at 302.8 nm is observed from the NIMS, LMI and HQ samples. Their same ZPL and LVM energy allow us undoubtedly to identify the presence of an identical center in the three hBN samples. Note that other deep defects are observed in Saint-Gobain and HQ samples. A ZPL is detected near 273 nm with a 71 meV LVM energy measured from 2 replica. This evidences another center in hBN crystals not yet referenced.

To conclude this part, the comparison of different hBN materials has revealed the richness of luminescence spectroscopy in the 200-400 nm range, which appears to be much more informative than Raman studies usually done. While the energy of intrinsic lines (S series) has been shown to be independent on the growth process, their intensity is found to be a relevant parameter for characterizing and benchmarking hBN crystals purity and crystallinity.

\subsubsection*{\qquad c - Comparative study of different hBN nanometer- thin layers\\\\}

\begin{figure}[ht]
 \centering
  \includegraphics[scale=0.6]{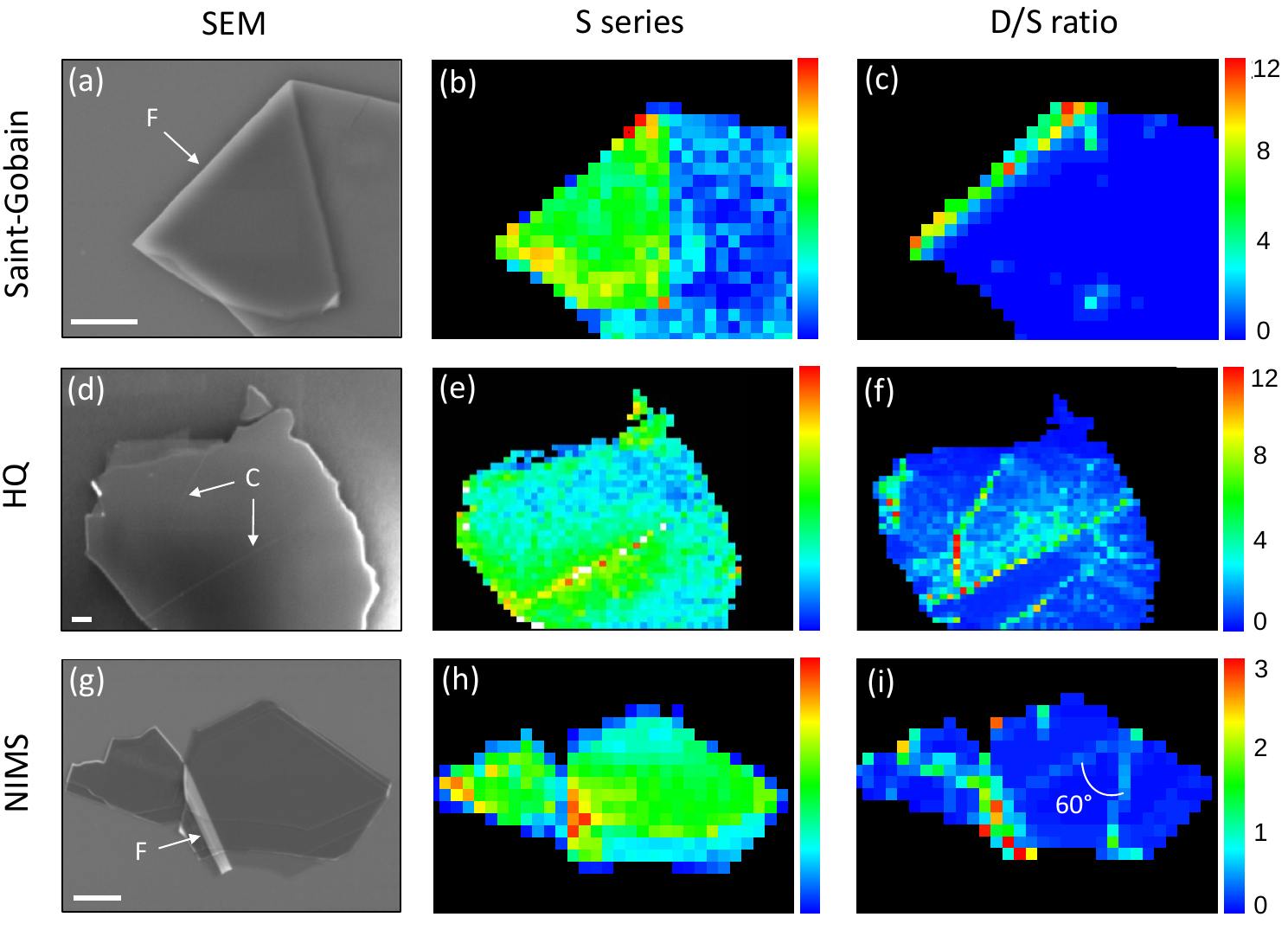}
  \caption{Defects investigation in three exfoliated hBN samples. SEM image, CL map of the S lines and of the D/S ratio have been recorded on flakes obtained from (a) (b) (c) the Saint-Gobain source, (d) (e) (f) a commercial HQ crystal and (g) (h) (i) the HPHT-grown single crystal from NIMS. T=10K. Scale bar: 2$\mu$m.}
  \label{F6}
\end{figure}

\qquad Beyond the characterization of bulk material sources, CL spectroscopy is also well suited to investigate the characteristics of nanometer-thin layers. We focus here on layers obtained by mechanical exfoliation from bulk sources as this is the material currently used for graphene engineering. Sheet thickness typically ranges within a few tens of nanometers and is too large to induce changes in the excitonic luminescence that we have evidenced  in \cite{Schue2016}. The key point here is to examine how to identify and localize the defects exfoliated sheets may contain. To this aim, we exploit an approach introduced in a previous work \cite{Pierret2014}, where we use the ratio of integrated intensities of defect-related (221-229 nm) and intrinsic (209-217 nm) excitonic series (D/S) to reveal the spatial distribution of structural defects in the material. We stress here that unlike the luminescence of deep centers described previously, the defect-related emissions of the D series in the NBE region are not systematically accompanied with a strong decrease of the intrinsic exciton luminescence. CL hyperspectral imaging (see details in Methods) were performed on various hBN flakes exfoliated from Saint-Gobain, HQ and NIMS sources. In figure \ref{F6}, cathodoluminescence maps of S lines and of the D/S ratio are shown together with the associated SEM image.

First, the flake presented in figure \ref{F6}(a) displays a fold, labeled F in the figure, often encountered in exfoliated 2D materials.  
The CL map of intrinsic S exciton recombinations depicted in figure \ref{F6}(b) clearly shows a homogeneous distribution in the two distinct parts of the flake. Considering the high energy of the incident electron beam, nanometer-thick hBN samples are transparent to the beam. As a result, the energy deposited in the flakes strongly depends on their thickness. In figure \ref{F6}(b), the increase in S intensity of the folded part is consistent with the increase of the generation factor $G$ expected for a thickness doubling (see \cite{Schue2016} for the detailed $G$ calculation in the case of transparent layers). By contrast, the D/S ratio allows to map a thickness independent signal as shown in figure \ref{F6}(c) and, as a result, is well homogeneous over the sample, except along the folding line where it is enhanced by a factor up to 12 (figure \ref{F6}(c)). Similar enhancement could be detected on several flakes meaning that folding generates defects in the atomic lattice which act as highly radiative recombination traps for excitons.

The same analysis was performed on the HQ sample (figure \ref{F6}(d-e-f)). In that example, intensity of the S lines is almost homogeneous over the flake in consistency with its constant thickness. Nonetheless, the D/S ratio map reveals inhomogeneities in the thin exfoliated layer. The D/S ratio increases along linear features which unveil the presence of structural defects. A part of the revealed defects was probably already present in the bulk source before exfoliation, as shown in figure \ref{F4} from the monochromatic CL image related to the D lines. 
However, the lines labeled C in figure \ref{F6}(d) appear localized along fracture lines in the basal plane also visible on the SEM image. They are crystallographically oriented in the easy cleavage planes perpendicular to the basal plane of hBN and might be the consequence of a too heavy load during the stamping process of the exfoliation procedure.

Finally, the NIMS exfoliated sample depicted in figure \ref{F6}(g) displays the most complicated situation, where thickness effects, folds, fractures and structural defects are encountered. Thickness steps are present on edges of the flake. Once again, the S intensity decreases on the thinnest parts while the D/S ratio remains almost constant. The D/S ratio is enhanced along the fold labeled F as reported previously and within the flake along lines with specific crystallographic orientations. (like the 60$^{\circ}$ angle indicated in figure \ref{F6}(i)). As a result of the presence of these defects the average value of the D/S ratio is generally above 0.8. To get more information on the nature of these defects, we performed TEM analysis of slabs cut along different crystallographic directions of the HPHT crystal as detailed now.\\

\subsubsection*{\qquad d - TEM analysis\\\\}

\qquad Two different slabs with thicknesses ranging from 40 to 70 nm were cut by Focused Ion Beam (FIB) in the HPHT single crystal source, tangentially (basal orientation [00.2] zone axis or plane view) and perpendicularly to the c - axis (prismatic orientation [11.0] zone axis or edge-on view).

\begin{figure}[ht]
 \centering
  \includegraphics[scale=0.95]{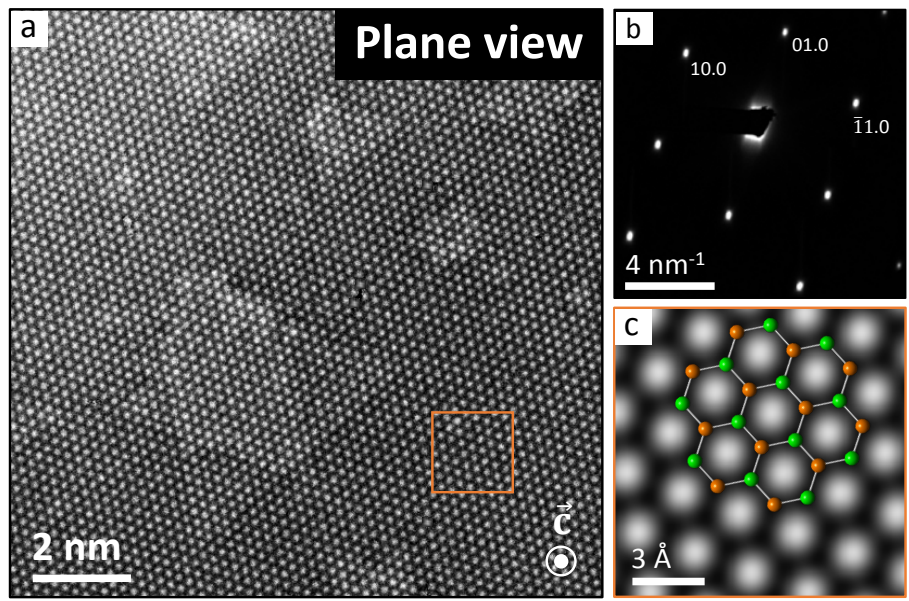}
  \caption{(a) HRTEM image of the hBN FIB sample in a plane view configuration and (b) the associated diffraction pattern exhibiting the typical hexagonal symmetry of the basal planes. (c) Fourier Filtered image of a zoom in image (a) with the B-N atoms positions. Note that the FIB sample is the same than the one investigated by CL imaging in figure \ref{F4}.}
  \label{F7}
\end{figure}

The plane view is well suited for visualizing the hexagonal BN lattice. The HRTEM image depicted in figure \ref{F7}(a) clearly evidences the typical features of the sp$^2$ network. Under the operating illumination conditions used, projected atomic potentials appear in dark while bright spots correspond to hexagon centers. In the Fourier Filtered (FF) image presented in figure \ref{F7}(c) the characteristic in-plane lattice parameter of 2.5~\AA~ is more accurately measured. Very characteristically, we observe large areas triangular in shape with a lighter contrast. Spatial extension of these areas ranges from atomic distances up to a few nanometers. As reported in several previous TEM works \cite{Alem2009,Jin2009,Meyer2009}, such contrast features are typical of vacancies aggregates in the h-BN network. These triangular areas observed here have two orientations, meaning that they are lying in different planes with the same edge termination (likely to be N terminated - edges according to \cite{Jin2009,Meyer2009}). Owing to the thickness of the slab and the illumination conditions used and operating at 80 kV, they cannot result from irradiation by the electron beam and should be considered as intrinsic growth defects in the single crystal.

\begin{figure}[ht]
 \centering
	\includegraphics[scale=0.55]{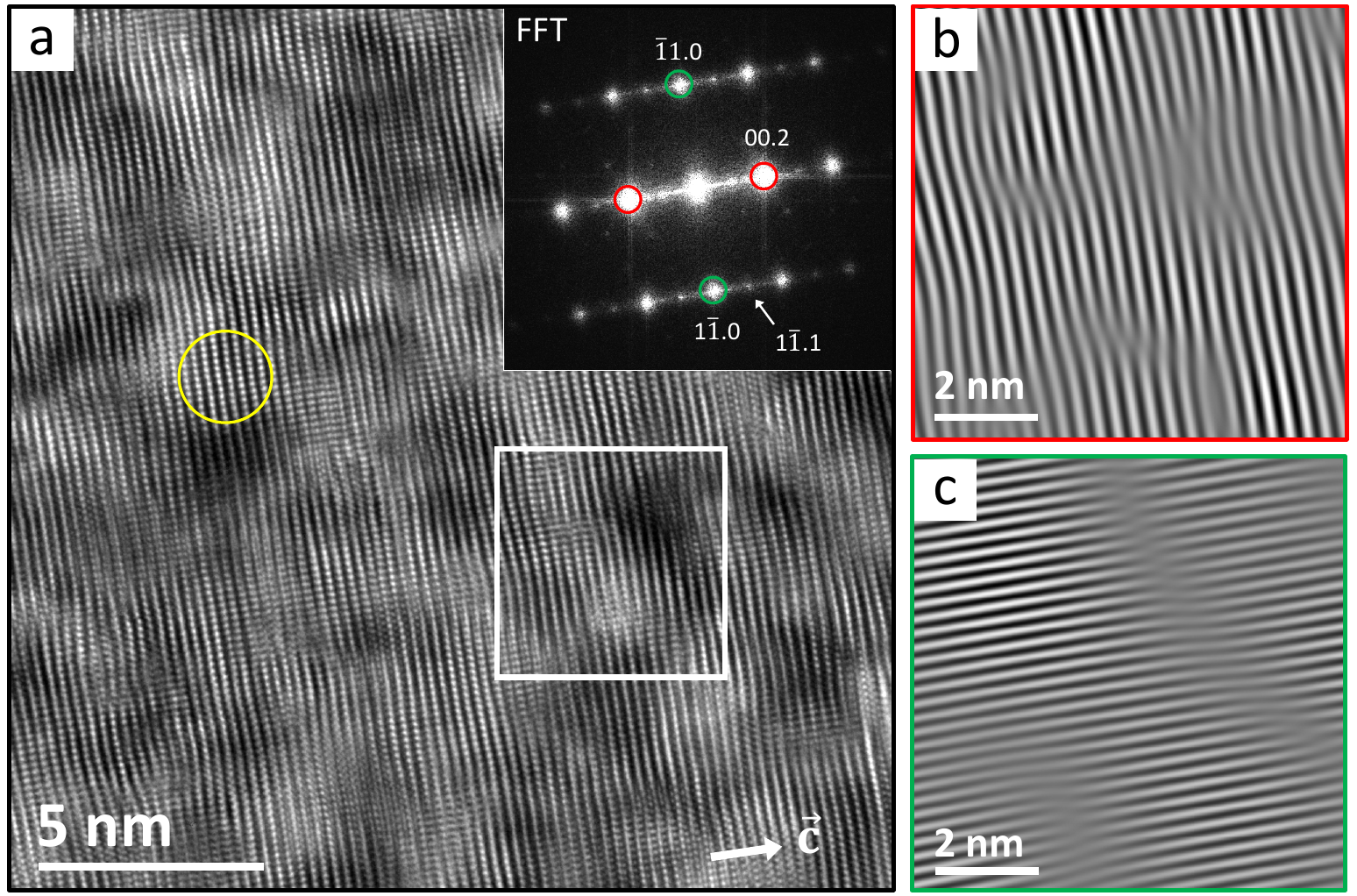}
	\caption{(a) HRTEM image of the hBN lamellae in edge view configuration; Inset: 2D Fast Fourier Transform (FFT) pattern of (a) using DigitalMicrograph software. (b) Fourier Filtered image of the area of (a) marked by the white square by selecting the $\{$00.2$\}$ and the $\{00.\bar{2}\}$ spots. (c) Fourier Filtered image of a reduced area (white square) obtained by selecting the $\{\bar{1}$1.0$\}$ and the $\{1\bar{1}.0\}$ spots.}
	\label{F8}
\end{figure}

The edge-on view (figure \ref{F8}(a)) provides a direct visualisation of the stacking sequence in hBN. As in plane-view images, projected atomic potentials exhibit a dark contrast. We can see in some regions (yellow circle) a perfect alignement of the BN atomic planes matching well with the expected AA' stacking sequence (see Supplementary Material for more details on the proper phase identification).

Additional information is however gained by inspecting this image at grazing incidence along a row perpendicular to the c axis. This reveals dots misalignement along the row due to structural perturbations within the basal (00.2) planes as already reported in the literature \cite{Huang1999}. These give rise to elongations of the diffraction spots along the [00.2] direction in the Fast Fourier Transform (FFT) pattern shown in inset.

Many structural perturbations are clearly more visible in the Fourier Filtered image of the region identified by the white square on the sample shown in figure \ref{F8}(b). It has been obtained by selecting the $\{$00.2$\}$ and the $\{00.\bar{2}\}$ reflections of the FFT pattern (red circles). In this filtered image, the fringes imaging (00.2) planes present discontinuities or distorsions due locally to the presence of an extra or missing fringe as at a dislocation line. Since the extension of the fringe in excess or in deficit can be up to a few nanometers, it is likely to link them to the triangular vacancies in the (00.2) planes observed in plane view on figure \ref{F7}.

But this is not all. Another kind of defect is revealed by considering now the Fourier Filtered image presented in figure \ref{F8}(c) obtained by selecting the $\{\bar{1}$1.0$\}$ and the $\{1\bar{1}.0\}$ reflections from the FFT pattern (green circles). In this filtered image, fringes, which are imaging (1$\bar{1}$.0) planes, can get shifted along the grey trace seen on the image. Such shifts are typically expected to arise if the stacking sequence of the (00.2) planes is locally changed from AA' to AB' or A'B by a simple translation (or a gliding) of the basal plane (see figure S2 in Supplementary Material).\\

In summary, close inspection of the sample at the atomistic scale by TEM has revealed the presence of two kinds of structural defects, namely large vacancies flakes and stacking faults, which could be responsible for the D line in cathodoluminescence spectra and for the intensity enhancements observed in the D/S mapping within the flake.

In conclusion of this part, inspection of thin exfoliated flakes from three different bulk sources reveals that the excitonic luminescence can be altered by different kinds of defects: folds, steps, modification in the layers stacking which can arise from the presence of large vacancies, change in the stacking sequence or dislocations. Furthermore, as this alteration produces an energy shift of the luminescence (D lines), this property can be exploited to get, in a non destructive way, the spatial distribution of these defects at nanometer scale.

\sectionred{Conclusion}

By comparing information one can get from Raman and cathodoluminescence spectroscopies on various samples, we have established a workable and efficient characterization procedure able to benchmark h-BN materials as bulk or as thin layers. Inspection of the full width of the 1366 cm$^{-1}$ Raman mode only provides a rough estimation of the sample quality at the macroscopic scale, usable for a preliminary benchmark. Interestingly, the energy of the interlayer shear mode detectable at very low frequency can be used to determine the number of layers, alternatively to usual AFM and optical imaging techniques. Nevertheless, as a whole, Raman spectroscopy cannot provide significant insight on the defects which may alter the properties of the material. By contrast, cathodoluminescence spectroscopy is shown to be very informative on the crystalline quality of the sample and the physical and chemical nature of the defects it may contain. This capability of the technique is due to the peculiar excitonic properties of h-BN (S lines) and the trapping of intrinsic excitons at structural defects (D lines). Using the intensity of the intrinsic emission lines, we benchmarked the purity and the crystallinity of four hBN samples. Moreover, we used the D/S intensity ratio to map the spatial distribution of structural defects in hBN exfoliated samples. 

This procedure has been shown to be fully operational for bulk samples and mechanically exfoliated flakes. Next step will be its application to the characterization of thinner layers synthesized by CVD techniques.

\sectionred{Methods}

\subsection*{\textbf{Sample preparation}}

\qquad The comparative study in Raman and CL has been done on four bulk sources obtained following different synthesis processes. A single crystal grown at high-pressure high-temperature (HPHT) provided by the NIMS \cite{Taniguchi2007} (NIMS sample) and a sample provided by the Laboratoire Multimat\'eriaux et Interfaces (LMI sample) obtained via a process coupling both Polymer Derived Ceramics (PDCs) and Spark Plasma Sintering (SPS) methods \cite{Yuan2016}. Then two commercially available materials were investigated : a hBN powder from the Saint-Gobain company Tr\`esBN-PUHP1108 (Saint-Gobain sample) and a crystal from the HQ Graphene company (HQ sample).

The BN sheets used for thickness determination and calibration measurements were exfoliated following the standard cleavage procedure from the Saint-Gobain powder. The FIB slabs investigated by TEM and CL were prepared from the HPHT-grown source.

\subsection*{\textbf{Experimental techniques}}

\qquad Thickness measurements by AFM were investigated with a Dimension 3100 scanning probe microscope (Brukers) operating in a tapping mode with commercial probes.

Conventional Micro-Raman characterization was performed using a Horiba Jobin Yvon Labram HR800 system using the 514.5 nm excitation wavelength of an Ar$^{+}$ laser with a power of 100 mW to excite the samples.

For low-frequency measurements, the laser plasma lines are removed using a narrow bandpass filter. The Rayleigh emission line is suppressed using three notch filters with a spectral bandwidth of 10 cm$^{-1}$. Argon gas is flowed over the sample to remove the low-frequency Raman modes from the air. We use a $\times$100 objective with NA = 0.8. A 1800 lines/mm grating enables us to have each pixel of the charge-coupled detector covering 0.57 cm$^{-1}$. A spectral resolution 0.7 cm$^{-1}$ is estimated from the width of the Rayleigh peak.

Transmission Electron Microscopy (TEM) observations were performed using a Zeiss Libra 200 MC equipped with an electrostatic CEOS monochromator, operating at 80kV and 200kV.

The cathodoluminescence experiments were conducted in (i) spectroscopic, (ii) imaging and (iii) spectral mapping modes using an optical system (HJY SA) installed on a JEOL7001F field-emission gun scanning electron microscope (SEM). The samples were mounted on a GATAN cryostat SEM-stage cooled down to 10K with a continuous flow of liquid helium. The temperature indicated in both CL spectra and the body text is always the temperature measured on the sample holder.
The CL emission is collected by a parabolic mirror and focused with mirror optics on the entrance slit of a 55cm-focal length monochromator. The all-mirror optics combined with a suitable choice of UV detectors and grating ensure a high spectral sensitivity down to 190 nm. A nitrogen cooled charge-coupled detector (CCD) camera is used to record the spectra in mode (i) and (iii). In the latter, also referred as the hyperspectral imaging acquisition mode, the focused electron beam is scanned step by step with the HJY CL Link drive unit and synchronized with the CCD camera to record one spectrum for each point.

The spectrometer is also equipped with an UV photomultiplier on the lateral side exit for fast monochromatic CL imaging (i.e., image of the luminescence at a given wavelength) in mode (ii).
The spectral response of the optical detection setup was measured from 200 to 400 nm using a deuterium lamp (LOT Oriel - Deuterium lamp DO544J - 30 W) of calibrated spectral irradiance. The system response for each detector/grating combination was then obtained following the procedure described in \cite{Pagel2000}.

\sectionred{Acknowledgments}

Catherine Journet-Gautier and Berang\`ere Toury-Pierre, from LMI, are warmly acknowledged for providing one of their PDCs samples. Authors thank T. Taniguchi and K. Watanabe from NIMS for providing us a reference HPHT crystal, David Troadec from IEMN for the FIB samples preparation and C. Vilar for technical help on cathodoluminescence-SEM setup. The research leading to these results has received funding from the European Union Seventh Framework Programme under grant agreement no. 696656 GrapheneCore1. We acknowledge funding by the French National Research Agency through Project No. ANR-14-CE08-0018.\\

\sectionred{References}

\bibliographystyle{unsrt}

\newpage

\title{\textbf{{\huge Supplementary Material}}}

\maketitle

\setcounter{figure}{0}
\renewcommand{\figurename}{Figure S}

\section*{{~\\ \Large Identification of the stacking sequence}}

Figure S1 presents a fine investigation of the BN stacking sequence through high resolution TEM imaging performed on the FIB slab analyzed in figure 8 in the body text. The region shown in figure S1a exhibits a well-organized structure of the BN basal planes as observed in the area identified by the orange square. The dot pattern, shown in the corresponding Fourier Filtered image (figure S1c) is fully consistent with the simulated image (figure S1e), assuming a AA' stacking sequence, which is the one expected for the h-BN structure. This result is even more confirmed by considering the electron diffraction pattern (figure S1b). Relative intensities of (1$\bar{1}$.0) and (1$\bar{1}$.1) reflections are matching the ones in the calculated pattern for the AA' stacking (figure S1d). Beside AA', the most energetically favorable stacking sequences AB and AB' would produce more intense (1$\bar{1}$.1) reflections with respect to the (1$\bar{1}$.0) ones (see calculated diffraction patterns of the possible stacking sequences in figure S2).\\

\begin{figure}[ht]
 \centering
  \includegraphics[scale=0.9]{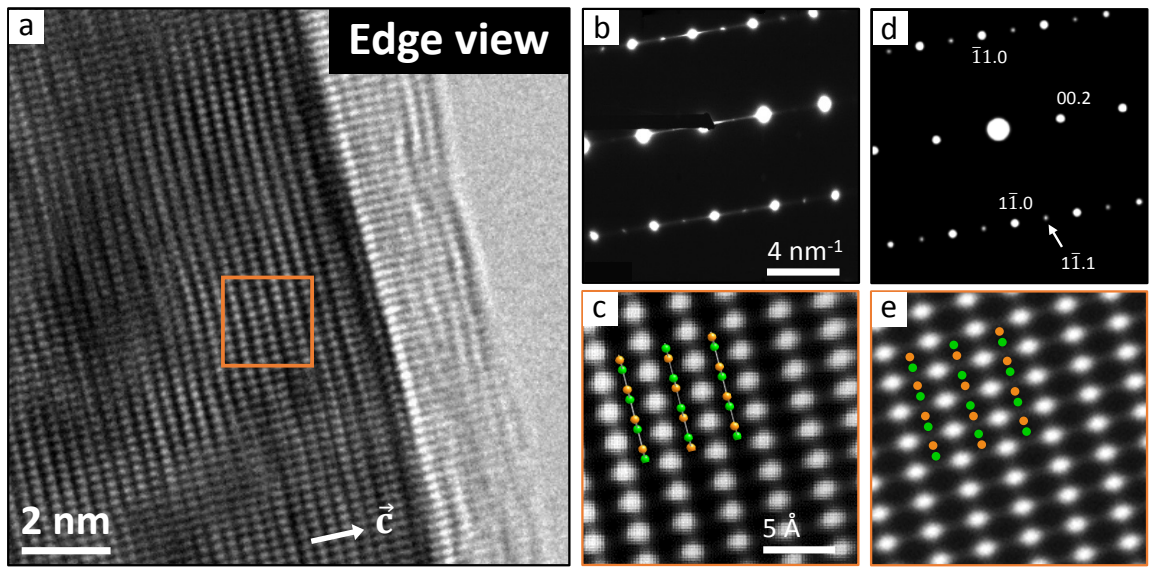}
  \caption{(a) HRTEM image of the hBN FIB sample in edge view configuration with associated (b) experimental diffraction pattern taken along the [11.0] zone axis (c) Fourier Filtered image of a zoom in (a) (orange square) with a scheme indicating B and N atom positions. (d) Simulated diffraction pattern and (e) HR image associated. Simulation was performed using JEMS software.}
  \label{FS1}
\end{figure}

\newpage

\section*{{~\\ \Large Simulated diffraction patterns as a function of the BN stacking sequence}}

\begin{figure}[ht]
 \centering
\includegraphics[scale=0.26]{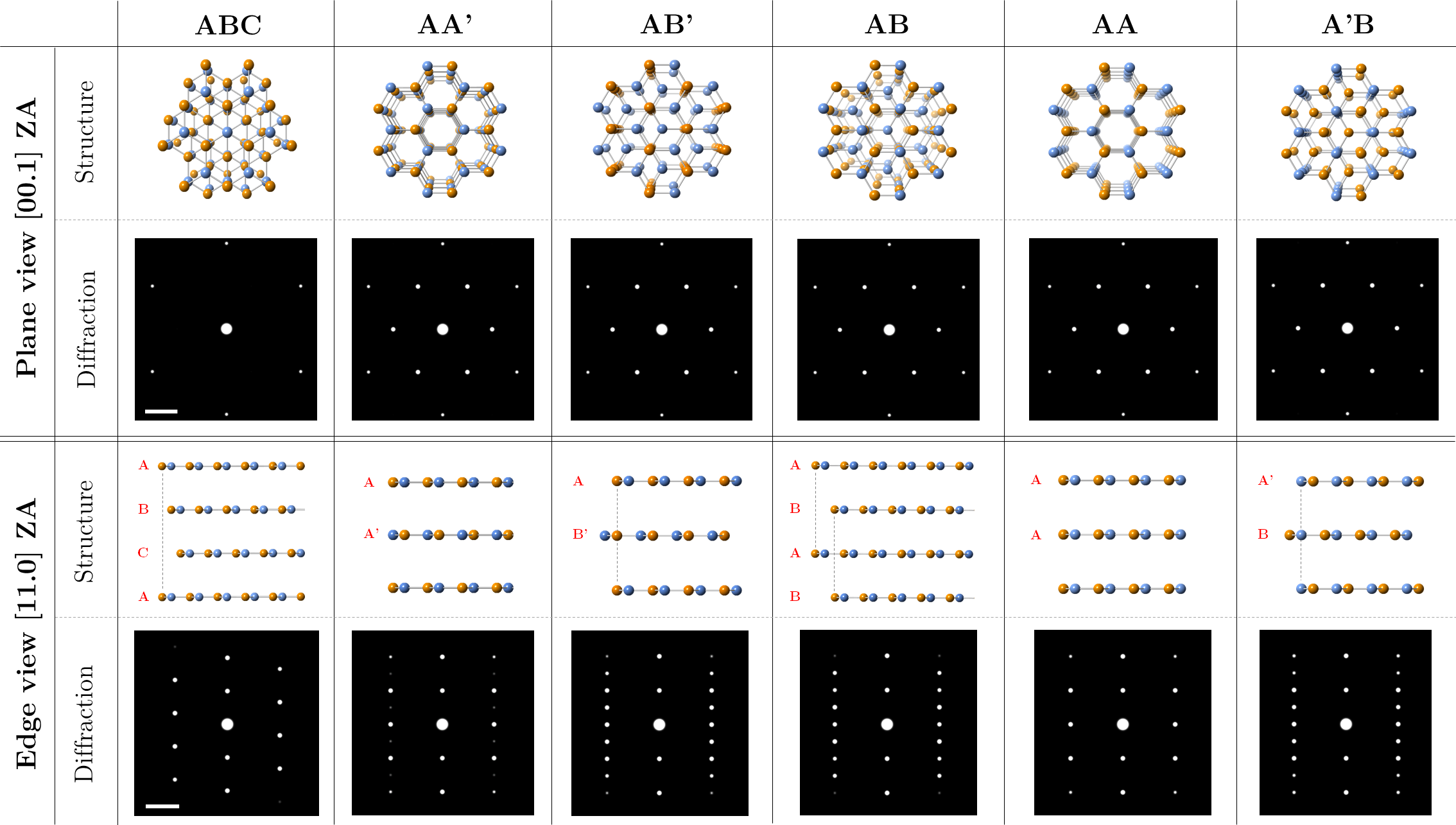}
\caption{Sketches of the possible stacking sequences of the BN sheets in plane view (upper part) and in edge view (lower part) and their corresponding calculated diffraction patterns.}
\label{FS2}
 \end{figure}

Figure S2 presents the different possible stacking sequences one can build from a BN sp$^{2}$ network. The energetic most favorable sequence is the AA' stacking, which well matches the experimental pattern discussed in the body text. The two other possible configurations are the AB' and A'B sequences through a simple translation or a gliding of the basal plane.  Their simulated diffraction patterns show a clear enhancement of the (1$\bar{1}$.1) and all other related spots unlike for the AA' sequence.

\end{document}